\documentclass[sn-mathphys,Numbered]{sn-jnl}

\usepackage{graphicx}%
\usepackage{multirow}%
\usepackage{amsmath,amssymb,amsfonts}%
\usepackage{amsthm}%
\usepackage{mathrsfs}%
\usepackage[title]{appendix}%
\usepackage{xcolor}%
\usepackage{textcomp}%
\usepackage{manyfoot}%
\usepackage{booktabs}%
\usepackage{algorithm}%
\usepackage{algorithmicx}%
\usepackage{algpseudocode}%

\usepackage{listings}%
\definecolor{codegreen}{rgb}{0,0.6,0}
\definecolor{codegray}{rgb}{0.5,0.5,0.5}
\definecolor{codepurple}{rgb}{0.58,0,0.82}
\definecolor{backcolour}{rgb}{0.95,0.95,0.92}
\lstdefinestyle{mystyle}{
  backgroundcolor=\color{backcolour}, commentstyle=\color{codegreen},
  keywordstyle=\color{magenta},
  numberstyle=\tiny\color{codegray},
  stringstyle=\color{codepurple},
  basicstyle=\ttfamily\footnotesize,
  breakatwhitespace=false,         
  breaklines=true,                 
  captionpos=t,                    
  keepspaces=true,                 
  numbers=left,                    
  numbersep=2pt,                  
  showspaces=false,                
  showstringspaces=false,
  showtabs=false,                  
  tabsize=1
}
\usepackage[font=small,labelfont=bf]{caption}

\usepackage{soul}

\usepackage{tcolorbox}
\newcommand{\find}[1]{
\begin{tcolorbox}[leftrule=1mm,toprule=0mm,bottomrule=0mm,left=1pt,right=2pt,top=2pt,bottom=2pt]
\em #1
\end{tcolorbox}
}

\theoremstyle{thmstyleone}%
\theoremstyle{thmstyletwo}%

\theoremstyle{thmstylethree}%

\begin{document}

\title[Article Title]{Adversarial Patch Generation for Automated Program Repair}


\author*[1]{\fnm{Abdulaziz} \sur{Alhefdhi}}\email{aa043@uowmail.edu.au}

\author[1]{\fnm{Hoa Khanh} \sur{Dam}}\email{hoa@uow.edu.au}

\author[2]{\fnm{Thanh} \sur{Le-Cong}}\email{congthanh.le@student.unimelb.edu.au}

\author[2]{\fnm{Bach} \sur{Le}}\email{bach.le@unimelb.edu.au}

\author[1]{\fnm{Aditya} \sur{Ghose}}\email{aditya@uow.edu.au}

\affil[1]{\orgdiv{School of Computing and Information Technology}, \orgname{University of Wollongong}, \orgaddress{\street{Northfields Ave.}, \city{Wollongong}, \postcode{2522}, \state{New South Wales}, \country{Australia}}}

\affil[2]{\orgdiv{School of Computing and Information Systems}, \orgname{The University of Melbourne}, \orgaddress{\street{Grattan St.}, \city{Parkville}, \postcode{3010}, \state{Victoria}, \country{Australia}}}



\abstract{Automated Program Repair has attracted significant research in recent years, leading to diverse techniques which focus on two main directions: search-based and semantic-based program repair. The former techniques often face challenges due to the vast search space, resulting in difficulties in identifying correct solutions, while the latter approaches are constrained by the capabilities of the underlying semantic analyser, limiting their scalability. In this paper, we propose NEVERMORE, a novel learning-based mechanism inspired by the adversarial nature of bugs and fixes. NEVERMORE is built upon the Generative Adversarial Networks architecture and trained on historical bug fixes to generate repairs that closely mimic human-produced fixes. Our empirical evaluation on 500 real-world bugs demonstrates the effectiveness of NEVERMORE in bug-fixing, generating repairs that match human fixes for 21.2\% of the examined bugs. Moreover, we evaluate NEVERMORE on the Defects4J dataset, where our approach generates repairs for 4 bugs that remained unresolved by state-of-the-art baselines. NEVERMORE also fixes another 8 bugs which were only resolved by a subset of these baselines. Finally, we conduct an in-depth analysis of the impact of input and training styles on NEVERMORE’s performance, revealing where the chosen style influences the model’s bug-fixing capabilities.}




\maketitle

\section{Introduction}

Software bugs are costly to detect and rectify~\cite{onoma1998regression,engstrom2010qualitative,bohme2014corebench}. 
Due to deadline-meeting demands, software programs are often delivered with known or unknown bugs~\cite{anvik2005coping}.
The number of bugs may be far more than the amount of human resources available to address them. It can take days or even years for software defects to be repaired~\cite{bohme2014corebench}. 
\emph{Automated} approaches to detect and rectify software bugs are, thus, of tremendous value.

Automated Program Repair (APR) is a growing research area focused on automatically fixing software bugs. In recent years, numerous APR techniques~\cite{weimer2009automatically,kim2013automatic,qi2015analysis,le2016history,xuan2016nopol,long2016automatic,mechtaev2016angelix,xiong2017precise,le2017s3,mahajan2018automated,liu2019tbar,chen2019sequencer,le2021usability, ye2022neural,xia2022less,ye2022selfapr} have been proposed, showing great promise in efficiently and effectively addressing real-world program defects. Notably, Facebook's introduction of GetaFix~\cite{getafix} in 2018, inspired by a recent research work~\cite{le2016history}, demonstrated the practical impact of APR research and its potential for practical adoption. GetaFix served as evidence that APR can revolutionise software development by automating bug fixes and enhancing the productivity of software developers. These advancements highlight the significance of the ongoing research in APR and its ability to reshape the industry.

Traditional APR techniques mainly fall into two main categories, namely \textbf{search-based} and \textbf{semantics-based} program repair. While semantics-based approaches use semantic analysis, such as static analysis and symbolic execution (e.g.,~\cite{mechtaev2016angelix,le2017s3}), search-based approaches often use syntactic analysis to generate and traverse syntactic search spaces such as genetic programming (e.g.,~\cite{weimer2009automatically}), pattern recognition (e.g.,~\cite{le2016history}), and machine learning (e.g.,~\cite{long2016automatic}). 
Semantics-based approaches are typically precise but limited by the capability of the underlying semantic analysers, (e.g., symbolic execution~\cite{mechtaev2016angelix, le2017s3}). Meanwhile, search-based approaches typically generate a large syntactic search space, rendering it difficult to navigate through to find correct solutions.

To tackle this challenge, recent advancements in APR have leveraged learning mechanisms to acquire the ability to rectify bugs based on historical bug fixes~\cite{le2016history, ye2022neural, xia2022less, ye2022selfapr}. These techniques, known as \textbf{learning-based} program repair, take advantage of the vast availability of software projects (e.g., 330+ million repositories on GitHub~\cite{about2023github}) to automatically identify and learn vulnerability patterns along with their corresponding patches. As a result, these techniques can efficiently propose solutions for newly discovered bugs.

Learning-based approaches in program repair primarily view the task as a translation process, aiming to transform buggy code into correct code. To achieve this goal, these approaches employ sequence-to-sequence models comprising an encoder and a decoder. The model is then trained on the historical bug-fixes data using cross-entropy loss, which measures the syntactic similarity between generated and ground truth patches for a given buggy code. 
In this paper, we introduce a novel approach that utilises the architecture of Generative Adversarial Networks (GANs)~\cite{goodfellow2014generative} for the automatic generation of bug fixes. GANs consists of two competing networks known as the \ul{generator} and \ul{discriminator}. Our approach incorporates an attention-based encoder-decoder network as the \emph{Patch Generator} and an LSTM network as the \emph{Patch Discriminator}. 
Similar to previous learning-based approaches, our GANs model learns from historical data comprising buggy code snippets and their corresponding fixes. However, our approach distinguishes itself by \textit{introducing a unique learning paradigm that establishes a competitive environment between the two neural networks: Patch Generator and Patch Discriminator}. This competitive nature leverages the adversarial characteristics of GANs, leading to a more effective patch generation.

We evaluated our approach, i.e., NEVERMORE (geNEratiVe advERsarial networks for autoMated prOgRam rEpair) on a dataset of 500 real-world bugs from the BigFix dataset~\cite{li2020dlfix}. Focusing on single-line bugs, our empirical evaluation demonstrated that NEVERMORE
generate fixes that are identical to human-written fixes for 21.2\% of the examined bugs. Based on this promising result, we further evaluate our approach on Defects4J, a well-known benchmark in program repair, and compare it with three well-known deep learning approaches that also target single-line bugs, including CocoNuT~\cite{lutellier2020coconut}, SequenceR~\cite{chen2019sequencer} and DLFix~\cite{li2020dlfix}. The experimental results showed that our approach can correctly fix 18 bugs, of which 4 bugs remain unsolved by the above state-of-the-art baselines while another 8 bugs were only fixed by a subset of these baselines. Finally, we conduct an in-depth analysis on the impact of input and training styles on NEVERMORE's performance, revealing that the chosen style significantly influences the model’s bug-fixing capabilities. We also demonstrate our detailed exploration procedure to demonstrate the importance of intermediate experiments in designing program repair techniques.

In summary, we make the following contributions:

\begin{itemize}
\item We introduce the concept of Generative Adversarial Networks (GANs) for Automated Program Repair. As discussed in Section \ref{sect:gan-potential}, our proposed framework enables the integration of various state-of-the-art patch generation methods into an adversarial environment, thereby enhancing their performance;
\item We propose NEVERMORE, a novel program repair technique that leverages the architecture of Generative Adversarial Networks (GANs) with a simple and light-weight patch generator and patch discriminator for fixing software bugs;
\item We empirically demonstrate the promising capabilities of our design in improving the performance of program repair techniques. Furthermore, we conduct a comprehensive analysis of the impact of input and training styles on the effectiveness of our approach, providing empirical insights for future research; 
\item We make our replication package, including experimentation codebase and results, publicly available for future research\footnote{\url{https://drive.google.com/drive/folders/18G061eW6axnElYUUkM7rGs0JqZsZoItH?usp=sharing}}.
\end{itemize}

\section{Background}

In this section, we briefly introduce Generative Adversarial Networks.




\subsection{Generative Adversarial Networks}
Many adversarial scenarios exist in life. For example, counterfeiters try to deposit fake money while banks try to detect it. The counterfeiters try to bring more realistic money each time while the banks try to improve their fake money detection power. These adversarial settings, whether initiated intentionally or unintentionally, can be exploited for learning and development purposes. The famous deep learning architecture, Generative Adversarial Networks (GANs), was introduced in 2014 by Goodfellow et al.~\cite{goodfellow2014generative} providing an adversarial setting within the neural ecosystem. The GANs architecture consists of two neural networks\footnote{The mechanism in which neural networks operate is inspired by how the neural networks in the human brain process information. An artificial neural network consists of \ul{layers} of \ul{nodes} connected to each other by \ul{weights}.}: a \textbf{generator} and a \textbf{discriminator}. The former generates new instances that resemble a given data population, while the latter aims to detect the authenticity of the generated instances. Both of them are connected and trained together in a zero-sum game style. The generator tries to ``fool'' the discriminator into thinking that the data it generates are ``real'', meaning that it is a data point from the given data population. The discriminator tries to distinguish between ``fake'' data produced by the generator and ``real'' data from the data population. The adversarial setting of GANs allows the two networks to learn from and, consequently, improve each other over time.

\section{Methodology}\label{sect:apg}

\begin{figure}[thb]
	\includegraphics[width=\linewidth]{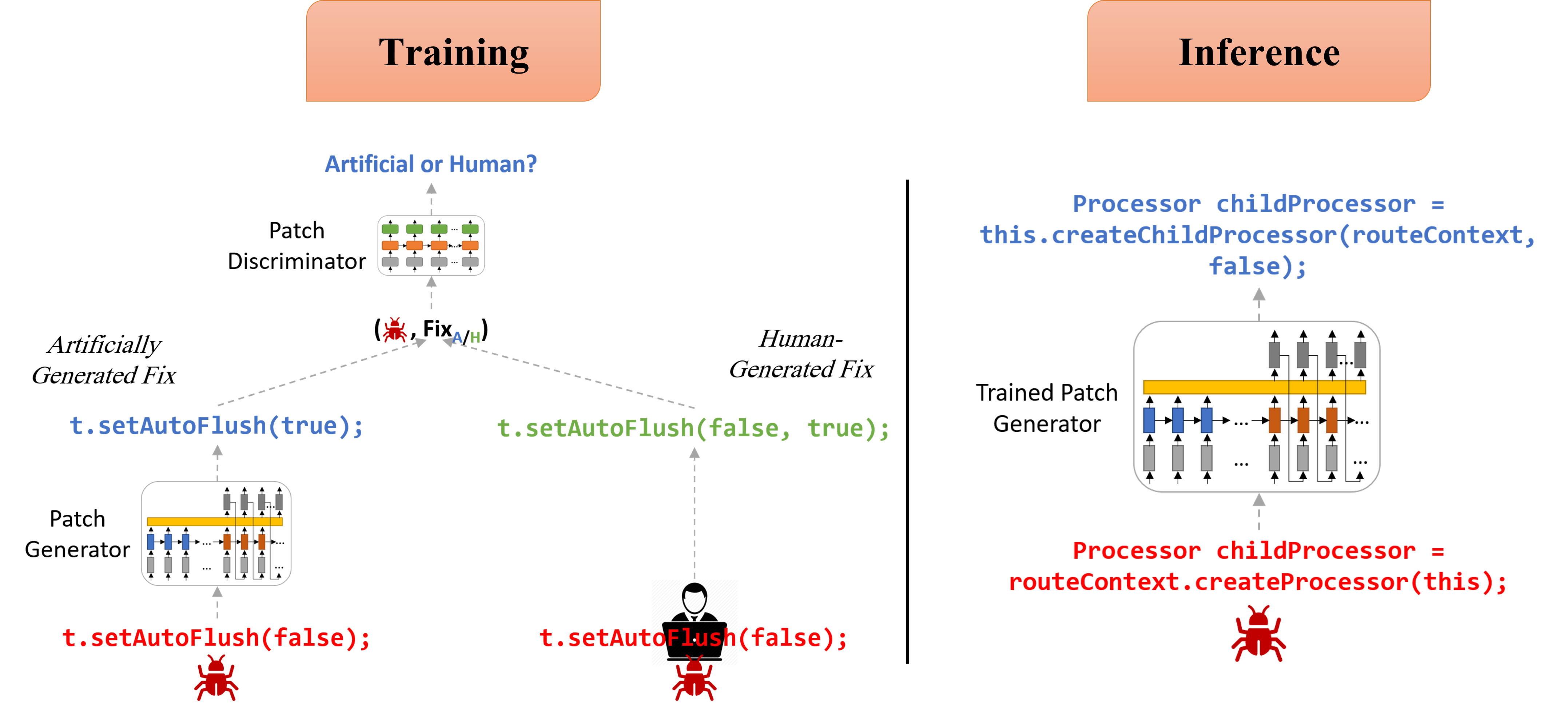}
	\caption{An overview of our approach. In the training phase, the \emph{Patch Discriminator} is used to help the \emph{Patch Generator} learn to imitate human fixes. The trained \emph{Patch Generator} is then used in the inference phase to generate patches for buggy code.}
	\label{fig:approach}
\end{figure}

\subsection{Motivating Example}
\begin{lstlisting}[language=Java, caption= A developer-written fixes extracted from the BigFix~\cite{li2020dlfix} dataset, label=fig:example]
- t.setAutoFlush(false);
+ t.setAutoFlush(false, true);
\end{lstlisting}

Listing~\ref{fig:example} illustrates a bug and its corresponding fix extracted from the BigFix dataset~\cite{li2020dlfix}. This bug was fixed by the developer by adding a new argument (\texttt{true}) to the method call.

In traditional deep learning, the training data typically serves as the exclusive source of information for neural networks to learn from. As a result, prior learning-based program repair approaches directly measure the syntactic distance between their generated patches with developer-written fixes to guide the learning algorithm. By contrast, our approach employs a GAN architecture, which consists of two neural networks: the Generator and Discriminator. These neural networks engage in a competitive environment, challenging and evaluating each other and, as a result, enhancing the performance of both networks~\cite{goodfellow2014generative}. 

\begin{lstlisting}[language=Java, caption= Example of a fix attempt generated by the Generator after few training iterations, label=fig:attempt_example]
- t.setAutoFlush(false);
+ t.setAutoFlush(true);
\end{lstlisting}

For example, suppose that after a few training iterations, the Generator produces the following fixing attempt for the bug in Listing~\ref{fig:example} (see Figure \ref{fig:approach}). In this attempt, the Generator replaces the passed argument from \texttt{false} to \texttt{true} instead of adding it as a new argument. In this example, the small syntactic difference between a fix attempt and a developer-written fix will challenge the traditional deep learning approaches. 

In contrast, in our scenario, the Generator and Discriminator are competing with each other in a zero-sum game. The Generator's main goal is to ``fool'' the Discriminator into thinking that it is a human by producing fixes that are as close as possible to the human-generated fixes. Meanwhile, the Discriminator's main mission is  to distinguish between human-generated and artificial fixes. 
If the Discriminator is able to distinguish between the human and artificial fixes, the Generator needs to step up its game and produce fixes that are more human-like. If the Discriminator fails to distinguish, it needs to step up its detection game. This adversarial mechanism allows the neural weights of both the Generator and Discriminator to be adjusted according to the gradient descent of the loss. In this example, the ultimate goal is for the Discriminator to keep detecting that \texttt{t.setAutoFlush(false, true);} is human-generated and any other fix is Generator-produced until the Generator is able to generate \texttt{t.setAutoFlush(false, true);}.

In summary, the GANs architecture allows not one but two neural networks to learn from the training data. In addition, they learn from and improve each other and, as a result, the Generator can be used to generate human-like fixes.

\subsection{Adversarial Patch Generation}\label{sect:app-des}
Figure~\ref{fig:approach} illustrates the overall framework of NEVERMORE. The workflow is inspired by the architecture of Generative Adversarial Networks (GAN), consisting of two main components: \emph{Patch Generator} and \emph{Patch Discriminator}. In this work, we adapt an extension of GANs, namely Conditional GANs (cGANs)~\cite{mirza2014conditional} since this architecture allows the \emph{Patch Generator}'s output, i.e., fixed code, to be conditioned on a certain input, i.e., buggy code. 


\textbf{Patch Generator.} Our \textit{Patch Generator} takes a buggy code fragment as input and utilizes its acquired knowledge from training data to generate a fix for the code. The goal of \textit{Patch Generator} is to generate patches that are as close as possible to human-written code. For example, given the buggy code:
\begin{lstlisting}
    t.setAutoFlush(false);
\end{lstlisting}
\textit{Patch Generator} will produce the following fix:
\begin{lstlisting}
    t.setAutoFlush(true);
\end{lstlisting}
in which the boolean parameter of the function \texttt{setAutoFlush} is changed from \texttt{false} to \texttt{true}. To achieve this goal, we employ an attention-based encoder-decoder network, which is commonly used for generation/translation tasks~\cite{bahdanau2014neural,sutskever2014sequence}. We use Long Short-Term Memory (LSTM) networks~\cite{hochreiter1997long} for the implementation of the encoder and decoder. 

\textbf{Patch Discriminator.} On the other hand \emph{Patch Discriminator} takes two 
pairs as input: (1) a pair of a buggy code and its human-generated patch and another pair of the same buggy code and (1) the patch that was generated by the \emph{Patch Generator}. For each pair, the \emph{Patch Discriminator} determines whether the fixed code is human-generated 
or artificially generated. For example, given the aforementioned fix generated by \textit{Patch Generator}, \emph{Patch Discriminator} will identify whether the change of the boolean parameter of the function \texttt{setAutoFlush} from \texttt{false} to \texttt{true} is artificially generated. To achieve this goal, we also employ Long Short-Term Memory (LSTM) networks~\cite{hochreiter1997long} but with sigmoid activation. Code fragment pairs are first tokenised into code sequences. We incorporate word embedding~\cite{gal2016theoretically} by adding an embedding layer at the beginning of both the generator and discriminator.
To aid our model in capturing the ground truth distribution, random noise is introduced by applying dropout~\cite{gal2016theoretically} to the network layers of the \emph{Patch Generator}.


Overall, the \emph{Patch Generator} and \emph{Patch Discriminator} will be trained in a competitive environment where one's loss is the other's win. Particularly, if the \emph{Patch Discriminator} correctly identifies the patch's source, the \emph{Patch Generator} tries to improve its future patches. If the \emph{Patch Generator} successfully fools the \emph{Patch Discriminator}, the \emph{Patch Discriminator} tries to improve its detection skills for future pairs.
This mechanism enables the \emph{Patch Generator} to eventually produce human-like fixes. It also improves the  \emph{Patch Discriminator} detection capabilities. Nevertheless, we only use the \emph{Patch Discriminator} during training in our study.

\subsection{Model Training}

We utilise the BigFix dataset~\cite{li2020dlfix} and the Defects4J benchmark~\cite{just2014defects4j} to train and test our model. 
Following previous studies (e.g.,~\cite{li2020dlfix,chen2019sequencer}), we focus on collecting single-line fixes. For more details regarding the data used in our study, refer to Section \ref{sect:emp_eval}.

During the training process, we employ the \emph{Patch Generator} to generate candidate repairs for each buggy code fragment in the training set. Subsequently, a \emph{Patch Discriminator} is employed to compare two pairs. The first pair consists of the faulty code (e.g., \texttt{\small t.setAutoFlush(false);}) and its corresponding human-fix obtained from the dataset (e.g., \texttt{\small t.setAutoFlush(false, true);}). The second pair consists of the same faulty code and the code generated by the \emph{Patch Generator} (e.g., \texttt{\small t.setAutoFlush(true);}). The \emph{Patch Discriminator} is trained to distinguish between the pair containing the human fix and the pair containing the artificially generated fix.

To optimize the model parameters, the \emph{Patch Discriminator} propagates the loss gradient, which quantifies its misclassification of artificial or human fixes, to the \emph{Patch Generator}. The \emph{Patch Generator} leverages this loss, combined with the loss computed from the difference between the generated fixes and the ground truth, to enhance its performance in producing fixes that resemble human-written ones, effectively attempting to `fool' the \emph{Patch Discriminator} by generating fixes that closely resemble human fixes.

\begin{sloppypar}After completing the training phase, we utilise the trained \emph{Patch Generator} for inference purposes, while discarding the \emph{Patch Discriminator}. The primary role of the discriminator in our study is to facilitate teaching and enhancement of the \emph{Patch Generator}. During the inference phase (refer to Figure \ref{fig:approach}), the trained \emph{Patch Generator} is presented with new instances of buggy code fragments (e.g., \texttt{\small Processor childProcessor = routeContext.createProcessor(this);} from the evaluation set). It then suggests patches for these buggy fragments (e.g., \texttt{\small Processor childProcessor = this.createChildProcessor(routeContext, false);}).\end{sloppypar}

\section{Empirical Evaluation}\label{sect:emp_eval}

In this section, we evaluate our approach of using GANs for bug-fixing. In Section \ref{never:rqs}, we present the research questions of this study. In Section \ref{Never:exp_setup}, we discuss our experimental setup, including the utilised datasets and performance measures. Sections \ref{never-rq1}, \ref{never-rq2}, and \ref{never-rq3} relay our answers to RQ1, RQ2, and RQ3, respectively.

\subsection{Research Questions}\label{never:rqs}
Our evaluation aims to answer the following research questions:

\noindent \textbf{RQ1:} \textit{Is NEVERMORE effective in fixing bugs?} This research question concerns NEVERMORE's ability to fix bugs in real-world projects. To ensure the quality of the data, we eliminate duplicate pairs. As a result, our final dataset contains 5,749 pairs of buggy and fixed code samples from the BigFix dataset (See Section \ref{Never:exp_setup_datasets}), with 500 pairs reserved for evaluation and the remaining used for training our model. 
We evaluate NEVERMORE by monitoring its identical fix rates (Section \ref{sec:metric}).

\noindent \textbf{RQ2:} \textit{How does NEVERMORE compare to existing deep learning approaches in APR?} This research question investigates the effectiveness of our approach in comparison to existing techniques. To answer this question, we employ NEVERMORE to fix bugs in the Defects4J dataset and then compare it to three state-of-the-art baselines, including SequenceR~\cite{chen2019sequencer}, DLFix~\cite{li2020dlfix}, and CoCoNuT~\cite{lutellier2020coconut}. These are some of the best candidate baselines for our comparative study because these models share the same scope as our study's: i) they are neural-based models, and ii) they focus on single-line fixes. Other existing models were not nominated for our comparative study due to potential unfair comparisons. For example, CURE~\cite{jiang2021cure} was excluded because it focuses on code change hunks rather than single-line fixes. 


As mentioned earlier, NEVERMORE employs an attention-based encoder-decoder as the Generator. While we could motivate our approach by comparing it to a similar Generator outside the GANs framework, i.e., without the Discriminator, we chose to compare it with SequenceR and DLFix, both of which incorporate complex encoder-decoder models within their approaches. This decision presented us with a more challenging task due to the advanced complexity of the competitors' models. Nonetheless, we will demonstrate that NEVERMORE successfully fixes bugs that remain unresolved by these existing approaches. Additionally, we include CoCoNuT as a third baseline to compare our approach with a distinct neural architecture, namely Convolutional Neural Networks (CNNs).

We use the BigFix dataset for training our approach and the Defects4J benchmark to evaluate and compare.

\noindent \textbf{RQ3:} \textit{How do input and training styles affect NEVERMORE's performance?}

To answer this research question, we conduct five different experiments gradually and systematically, where each experiment is built based on the insights gained from the previous ones. We train NEVERMORE on the BigFix dataset and evaluate the trained models on Defects4J. The first three experiments study the relation between the redundancy and the level of abstraction of the training data and their impact on our model's performance. The last two experiments study the different context levels of the bug and their impact on our model's performance.

\subsection{Experimental Setup}\label{Never:exp_setup}

\subsubsection{Datasets}\label{Never:exp_setup_datasets}
\label{sec:dataset} To evaluate the effectiveness of our approaches we use two well-known datasets: BigFix and Defects4J. 

\begin{itemize}
    \item \textit{BigFix:} BigFix is built by Li et al. and used to train and test DLFix~\cite{li2020dlfix}. BigFix is built from a public bug detection dataset~\cite{li2019improving} consisting of +1.8 million buggy Java methods and their fixed version. Li et al. target one-line bugs and set up several filters to build up BigFix, including i) Filtering bugs that are inside a method, ii) Filtering out commented bugs, and iii) filtering one-hunk bugs. As a result, BigFix consists of +20K buggy-fixed pairs of source code.
    \item \textit{Defects4J:} The Defects4J dataset~\cite{just2014defects4j} has become a widely-used benchmark for software engineering research and program repair tasks. Defects4J is a collection of and a supporting infrastructure of real reproducible bugs from well-known open-source projects. While Defects4J at the time of writing this thesis contains 835 bugs (version 2.0.0\footnote{Accessible through \url{https://github.com/rjust/defects4j}.}), we used Defects4J version 1.2.0, containing 395 bugs, to be consistent with the version used in the baselines of this study even though NEVERMORE has the potential to fix more bugs had we used the latest Defects4J version.
\end{itemize}
\subsubsection{Performance Measure}\label{sec:metric}
We estimate the effectiveness of NEVERMORE by monitoring its \emph{identical fix rate}. Simply put, we report the number and percentage of model-generated fixes that exactly match the fixes from the ground-truth. We will report and discuss some model-generated fixes that are slightly different from the ground-truth yet provide insightful information.

\subsection{NEVERMORE's Performance}\label{never-rq1} 
\textbf{Answering RQ1:} Our empirical evaluation of 500 bugs extracted from BigFix demonstrates that NEVERMORE can effectively fix real-world bugs. Particularly, our model generates 106 fixes (21.2\%) that are identical to the human fixes (i.e., the ground truth). 

To further understand the repair capabilities of NEVERMORE, we also perform a qualitative analysis of the patches generated by NEVERMORE. To demonstrate the findings from our analysis, we provide the following examples. Listing~\ref{fig:example1} presents an example of an identical fix generated by our model. 
\begin{lstlisting}[language=Java, caption= Example of an identical fix generated by NEVERMORE, label=fig:example1]
// Human-written fix
- return new HiveQueryResultSet.Builder()
+ return new HiveQueryResultSet.Builder(null)

// NEVERMORE-generated fix
- return new HiveQueryResultSet.Builder()
+ return new HiveQueryResultSet.Builder(null)
\end{lstlisting}
The example highlights NEVERMORE's capability to generate fixes that perfectly match human-written fixes. In this example, our model correctly adds the new argument (\texttt{\small null}). 

Besides providing identical fixes for real-world bugs, we also found that NEVERMORE also provides partial fixes which can provide useful hints for software developers.
Listing~\ref{fig:example2} demonstrates an example in which our model is capable of generating a partial solution for a bug in the BigFix dataset.  
\begin{lstlisting}[language=Java, caption= Example of an partial fix generated by NEVERMORE, label=fig:example2]
// Human-written fix
- detectDeadlock(e, "unlock");
+ detectDeadlock(dbConn, e, "unlock" );

// NEVERMORE-generated fix
- detectDeadlock(e, "unlock");
+ detectDeadlock(dbConn, e, +++e);
\end{lstlisting}
Particularly, our model was able  to correctly suggest the addition of the argument \texttt{\small dbConn} to fix the bug, but it changes the last argument (from \texttt{\small unlock} to \texttt{\small +++e}), rendering a syntactic error. These errors can be detected by involving a program analysis filter, such as a syntax checker, which checks for the syntactic correctness of the suggested patches. This filtering functionality can be added as a post-processing step of the \emph{Patch Generator} to ensure that candidate patches with syntax errors are replaced with syntactically correct patches. Alternatively, this functionality can be added to the \emph{Patch Discriminator} to augment its capability of detecting patches with syntax errors. The \emph{Patch Generator} will indirectly benefit from that due to the adversarial learning setting of the approach.

Listing~\ref{fig:example3} demonstrates another partial fix generated by NEVERMORE:
\begin{lstlisting}[language=Java, caption= Example of a partial fix generated by NEVERMORE, label=fig:example3]
// Human-written fix
- tilities.clearWorkMap(); 
+ tilities.clearWorkMap(jconf); 

// NEVERMORE-generated fix
- tilities.clearWorkMap(); 
+ tilities.clearWorkMap(jc); 
\end{lstlisting}
This example demonstrates a case where the generated patch is similar to the human fix in suggesting the adding of an argument to the method \texttt{\small clearWorkMap()}. However, they differ in terms of the variables used: our model suggested using the variable \texttt{\small jc}, while the human developer used \texttt{\small jconf}. If \texttt{\small jc} does not exist in the original code file, the filtering functionality described above can detect and replace this candidate patch. However, if \texttt{\small jc} exists, this presents two possibilities. The first possibility is that the model-generated patch is better than, or as good as, the human-generated patch. This possibility inspires building test suites to examine the behaviour of the generated fixes. The second possibility is that the model-generated patch is not preferable. This provides another improvement direction for our work in terms of widening the input scope to capture the context of the surrounding code, including definition-use dependencies of variables. 

We also found that not only is NEVERMORE useful for simple bugs, it also is for complex ones. Listing~\ref{fig:example4} demonstrates an example in which our model is capable of generating complex fixes, which involve multiple correct changes. 
\begin{lstlisting}[language=Java, caption= Example of a complex fix generated by NEVERMORE, label=fig:example4]
// Human-written fix
- Processor childProcessor = routeContext.createProcessor(this);
+ Processor childProcessor = this.createChildProcessor(routeContext, true);

// NEVERMORE-generated fix
- Processor childProcessor = routeContext.createProcessor(this);
+ Processor childProcessor = this.createChildProcessor(routeContext, false);
\end{lstlisting}
In this example, NEVERMORE added \texttt{\small this.} in front of the method name, providing the correct method \texttt{\small createChildProcessor} instead of \texttt{\small createProcessor}, and having the correct number of arguments passed to the method with the correct first argument \texttt{\small routeContext}. Compared to state-of-the-art repair techniques such as~\cite{weimer2009automatically,le2016history,le2017s3}, this result is impressive as the fix involves multiple actions/mutations altogether. The model however suggests the incorrect value of the last argument (\texttt{\small false} instead of \texttt{\small true}) while it still learns to pass a boolean argument. Although this looks simple, it requires the capability of understanding the semantics of the code and its intended behaviour. Semantics-based approaches (e.g., ~\cite{mechtaev2016angelix, le2017s3}) use underlying semantic analysers (e.g., symbolic execution) to provide this capability. Our approach can be equipped with a semantics-based tool to explore its syntactic and semantic reasoning potential.

\find{
\textbf{Answers to RQ1:} Overall, NEVERMORE can effectively fix bugs, generating 106 identical fixes out of 500 bugs (21.2\%). Our deeper analysis also demonstrates the capabilities of NEVERMORE in suggesting human-like and complex fixes.
}


\subsection{Comparison to existing baselines}\label{never-rq2} 

\textbf{Answering RQ2:} We conduct a comparative study to evaluate our approach with three well-known deep learning approaches in the APR space, including:

\begin{itemize}
    \item \textit{SequenceR:} SequenceR, proposed by Chen et al. \cite{chen2019sequencer}, is an end-to-end program repair model based on sequence-to-sequence learning. To address the out-of-vocabulary problem, which is more prevalent in source code compared to natural language, SequenceR utilizes the copy mechanism \cite{see2017get}.
    
    \item \textit{DLFix:} DLFix, introduced by Li et al. \cite{li2020dlfix}, is a context-based code transformation learning model. It consists of a two-layer tree-based architecture, where the first layer learns from the Abstract Syntax Trees (ASTs) of the bug's context, and the second layer learns the tree transformation from the bug to the fix.
    
    \item \textit{CocoNuT:} CoCoNuT takes a different approach to program repair by employing Convolutional Neural Networks (CNNs). Lutellier et al. \cite{lutellier2020coconut} use ensemble learning within CoCoNuT and leverage the strengths of CNNs in extracting hierarchical features and modelling source code at various levels of granularity.
    
\end{itemize}

Table \ref{tbl:baseline-comparison} lists the 18 Defects4J bugs that are successfully treated by NEVERMORE. The table also shows where in these bugs the baseline approaches failed and where they succeeded.

The table shows that CoCoNuT has more fixed Defects4J bugs in common with our approach than SequenceR and DLFix. NEVERMORE still addresses bugs that CoCoNuT fails to fix. It is worth noting that CoCoNuT is trained on a large-scale dataset and demands significant computational resources and time. We believe that if the same circumstances are made available to NEVERMORE (which could be a potential future research direction), it will probably be capable of fixing many more bugs, including those that both the baselines can and cannot resolve.

Listings \ref{fig:chart8}, \ref{fig:math46}, \ref{fig:math49}, and \ref{fig:mockito13} shows the Defects4J bugs that were uniquely fixed by NEVERMORE. In chart 8, NEVERMORE replaces the wrong argument with the right argument \texttt{zone}. We notice that in Math 49 and Mockito 13, NEVERMORE demonstrates its superiority in deleting problematic code. For Math 46, some can argue that it is a replacement task while others can say it is a deletion task. However, we argue that NEVERMORE performed most impressively with Chart 8 where not only did it replace the wrong argument with the correct one (\texttt{zone}), it brought the correct argument from outside of the bug's vocabulary. Due to space limitations, we refer the reader to the Defects4J dataset for more information on its bugs.

\begin{table}[htb]
    \centering
	\caption{Comparing NEVERMORE with three DL-based approaches that target single-line fixes, namely SequenceR, DLFix, and CoCoNuT, on \emph{defects4j}. The table shows which DL-based approach, if any, can fix the same bugs as the ones that are successfully fixed by NEVERMORE.}
	\label{tbl:baseline-comparison}
	\begin{tabular}{||l||c|c|c|c||}
		\hline \hline
		\textbf{Bug ID} & \textbf{SequenceR} & \textbf{DLFix} & \textbf{CoCoNuT} & \textbf{NEVERMORE} \\ \hline \hline
		Chart 1 & Fixed & Fixed & Fixed & Fixed \\ \hline
		Chart 8 & - & - & - & Fixed \\ \hline
		Chart 11 & Fixed & Fixed & Fixed & Fixed \\ \hline
        Closure 18 & Fixed & - & Fixed & Fixed \\ \hline
        Closure 62 & - & Fixed & - & Fixed \\ \hline
        Closure 63 & - & Fixed & - & Fixed \\ \hline
        Closure 70 & - & - & Fixed & Fixed \\ \hline
        Closure 73 & Fixed & Fixed & Fixed & Fixed \\ \hline
        Closure 86 & Fixed & Fixed & Fixed & Fixed \\ \hline
        Lang 29 & - & - & Fixed & Fixed \\ \hline
        Math 22 & - & - & Fixed & Fixed \\ \hline
        Math 46 & - & - & - & Fixed \\ \hline
        Math 49 & - & - & - & Fixed \\ \hline
        Math 57 & Fixed & Fixed & Fixed & Fixed \\ \hline
        Math 65 & - & - & Fixed & Fixed \\ \hline
        Math 82 & Fixed & Fixed & Fixed & Fixed \\ \hline
        Mockito 13 & - & - & - & Fixed \\ \hline
        Time 19 & - & - & Fixed & Fixed \\ \hline \hline
	\end{tabular}
\end{table}

\begin{lstlisting}[language=Java, caption= Chart 8 from Defects4J, label=fig:chart8]
- this(time, RegularTimePeriod.DEFAULT_TIME_ZONE, Locale.getDefault());
+ this(time, zone, Locale.getDefault());
\end{lstlisting}

\begin{lstlisting}[language=Java, caption= Math 46 from Defects4J, label=fig:math46]
- return isZero ? NaN : INF;
+ return NaN;
\end{lstlisting}

\begin{lstlisting}[language=Java, caption= Math 49 from Defects4J, label=fig:math49]
- Iterator iter = res.entries.iterator();
+ Iterator iter = entries.iterator();
\end{lstlisting}

\begin{lstlisting}[language=Java, caption= Mockito 13 from Defects4J, label=fig:mockito13]
- if (verificationMode instanceof MockAwareVerificationMode && ((MockAwareVerificationMode) verificationMode).getMock() == invocation.getMock()) {
+ if (((MockAwareVerificationMode) verificationMode).getMock() == invocation.getMock()) {
\end{lstlisting}

\find{
\textbf{Answers to RQ2:} Our proposed approach successfully fixed 18 bugs in the Defects4J dataset, out of which 4 were exclusively resolved by NEVERMORE, while another 8 bugs remained unresolved by at least one of the state-of-the-art baselines.
}


\subsection{Impact of input and training styles}\label{never-rq3} 

\textbf{Answering RQ3:} The following is the chronological order of the experiments, along with their labels and descriptions:

\begin{enumerate}
    \item \textit{dups:} In model training, it is often encountered that the training data contains duplicate entries. In this experiment, we start with taking the training data as it is after processing \textbf{without removing} the duplicated data points. The purpose of this experiment is to investigate the impact of these duplicated data points on the performance of NEVERMORE. For this experiment, we utilise 11,237 BigFix data points (i.e., bug-fix pairs) to train NEVERMORE.

    \vspace{0mm}
    
    \item \textit{no-dups:} In this experiment, we focus on removing the duplicated entries from the training data. Although the obvious advantage of this step is reduced training time, we also examine its potential impact on NEVERMORE's bug-fixing capabilities within the benchmark. It's important to note that we apply the removal of duplicated data \textbf{after} applying the local abstraction technique. By doing so, we not only eliminate duplicated concrete code entries but also eliminate duplicated "abstracted" entries. Consequently, the dataset for NEVERMORE's training is reduced to 6,830 BigFix pairs as a result of this procedure. This experiment allows us to assess the impact of data duplicate removal on NEVERMORE's performance in addressing the benchmark's bugs.
    
    \vspace{0mm}

    \item \textit{partial-dups:} In this experiment, we apply data duplicate removal \textbf{before} applying the local abstraction technique. This means that duplicated concrete code entries are eliminated, while duplicated abstracted entries are retained. As a result, we have 8,859 BigFix pairs available for NEVERMORE's training. It's worth noting that this experiment provides NEVERMORE with a larger training dataset compared to the "no-dups" experiment but a smaller dataset compared to the "dups" experiment. This experiment is motivated by the observation that the "no-dups" experiment resulted in fewer Defects4J bugs being fixed compared to the "dups" experiment. We aim to investigate if the differences in bug-fixing performance were affected by the number of training data points. Subsequently, we will demonstrate that this intermediate experiment was indeed justified, as it enabled NEVERMORE to address more Defects4J bugs compared to both the "dups" and "no-dups" experiments.

    \vspace{0mm}

    \item \textit{3lc:} In this experiment, denoted as ``3lc,'' we include the context of the bug by incorporating \textbf{3 lines of code} surrounding the buggy line. The purpose of adopting this approach is to assess the degree to which NEVERMORE learns from the contextual information associated with bugs. As we will detail how the previous experiments showed that the \textit{partial-dups} is the most effective training data, we adopt this style for the training data going forward. As a result, NEVERMORE is trained on a dataset consisting of 8,913 BigFix bug-fix pairs, each accompanied by their respective 3-line context. 
    

    
    \vspace{0mm}

    \item \textit{mc:} As we will detail, the previous experiment, namely \textit{3lc}, exhibited a promising improvement by incorporating the bug context into the model input. Therefore, we decided to conduct further experimentation concerning the bug context, which is substantiated by our subsequent evaluation. In this new experiment, denoted as \textit{mc}, we extend the scope of bug context by including the entire \textbf{method} containing the bug within the model input. This expanded context provides a more comprehensive representation of the bug's surroundings. By including the entire method, we aim to explore the potential benefits derived from incorporating a broader context.
    For this experiment, we utilise a dataset comprising 9,240 BigFix pairs to train NEVERMORE. The incorporation of the entire method as part of the model input allows us to investigate the impact of a more extensive bug context on NEVERMORE's bug-fixing performance.
    
\end{enumerate}

Table \ref{tbl:indepth-analysis} shows the Defects4J bugs that were successfully fixed by every experiment. NEVERMORE fixes 7, 6, 8, 9, and 13 Defects4J bugs in the \textit{dups}, \textit{no-dups}, \textit{partial-dups}, \textit{3lc}, and \textit{mc}, respectively. The highest number of fixed bugs was achieved in the \textit{mc} experiment. We attribute that to the inclusion of the entire method containing the buggy line in the model input compared to the other experiments that either do not include the context or include only 3 lines of context.

Five Defects4J bugs were uniquely fixed by one of the five experiments. They are as follows: Chart 8 (Listing \ref{fig:chart8}) was solely fixed in \textit{3lc}. Chart 11 (Listing \ref{fig:chart11}), Closure 70 (Listing \ref{fig:closure70}), and Math 65 (Listing \ref{fig:math65}) were solely fixed in \textit{mc}. Finally, Mockito 13 (Listing \ref{fig:mockito13}) was solely fixed in \textit{partial-dups}.

We can observe from these findings that the more context included in the model input the better the model performs as the \textit{mc} experiment both fixes more Defects4J bugs in general as well as unique bugs than all the other experiments that either include less or no context. However, we can also observe that experiments with less context involvement have a certain advantage. The 2 uniquely fixed Defects4J bugs with less context (i.e., Chart 8 in \textit{3lc} and Mockito 13 in \textit{partial-dups}) were also uniquely fixed by NEVERMORE, unlike unique \textit{mc} fixes that are also fixed by one or more of the baselines. We argue that there is a certain level of dependency on the context in the baseline models that may influence their focus on the bug itself. Therefore, a desirable model is one that finds the perfect balance between learning from the bug itself and the surrounding context. Due to space limitations, refer to the Defects4J dataset to learn more about the bug contexts.

\begin{table}[htb]
    \centering
	\caption{In-depth analysis of the performance of NEVERMORE. We build five experimental setups as follows. \textbf{dups}: duplicated training data points kept. \textbf{no-dups}: applying local abstraction AFTER removing duplicated training data points (i.e. removing duplicated patterns). \textbf{partial-dups}: applying local abstraction BEFORE removing duplicated training data points (i.e., removing duplicated codes). \textbf{3lc}: including three lines of context before AND after the buggy line. \textbf{mlc}: including the entire method of the buggy line.}
	\label{tbl:indepth-analysis}
	\begin{tabular}{||l||c|c|c|c|c||}
		\hline \hline
		\textbf{Bug ID} & \textbf{dups} & \textbf{no-dups} & \textbf{partial-dups} & \textbf{3lc} & \textbf{mc} \\ \hline \hline
		Chart 1 & - & Fixed & - & - & Fixed \\ \hline
		Chart 8 & - & - & - & Fixed & - \\ \hline
		Chart 11 & - & - & - & - & Fixed \\ \hline
        Closure 18 & Fixed & Fixed & - & - & Fixed \\ \hline
        Closure 62 & - & - & - & Fixed & Fixed \\ \hline
        Closure 63 & - & - & - & Fixed & Fixed \\ \hline
        Closure 70 & - & - & - & - & Fixed \\ \hline
        Closure 73 & - & Fixed & Fixed & Fixed & - \\ \hline
        Closure 86 & Fixed & - & Fixed & Fixed & Fixed \\ \hline
        Lang 29 & - & - & - & Fixed & Fixed \\ \hline
        Math 22 & Fixed & - & Fixed & Fixed & Fixed \\ \hline
        Math 46 & Fixed & - & - & - & Fixed \\ \hline
        Math 49 & - & Fixed & Fixed & Fixed & - \\ \hline
        Math 57 & Fixed & - & Fixed & - & - \\ \hline
        Math 65 & - & - & - & - & Fixed \\ \hline
        Math 82 & Fixed & Fixed & Fixed & - & Fixed \\ \hline
        Mockito 13 & - & - & Fixed & - & - \\ \hline
        Time 19 & Fixed & Fixed & Fixed & Fixed & Fixed \\ \hline \hline
        \textbf{Total} & 7 & 6 & 8 & 9 & 13 \\ \hline \hline
	\end{tabular}
\end{table}

\begin{lstlisting}[language=Java, caption= Chart 11 from Defects4J, label=fig:chart11]
- PathIterator iterator2 = p1.getPathIterator(null);
+ PathIterator iterator2 = p2.getPathIterator(null);
\end{lstlisting}

\begin{lstlisting}[language=Java, caption= Closure 70 from Defects4J, label=fig:closure70]
- jsDocParameter.getJSType(), true);
+ jsDocParameter.getJSType(), false);
\end{lstlisting}

\begin{lstlisting}[language=Java, caption= Math 65 from Defects4J, label=fig:math65]
- chiSquare += residual * residual / residualsWeights[i];
+ chiSquare += residual * residual * residualsWeights[i];
\end{lstlisting}

\find{
\textbf{Answers to RQ3:}\label{Never:rq3} Duplication removal significantly enhances the quality of deep learning models. However, our investigations indicate that local abstractions do not contribute to this improvement. Our experiments also suggest that incorporating the context, especially the method-level context, of the bug in the model input, improves models' performance. 
}



\section{Discussion}
\subsection{The Potential of GANs for APR}\label{sect:gan-potential}

Breaking down the name GANs, i.e. Generative Adversarial Networks, it provides three fundamentals: \textbf{i) generating} a desired product, \textbf{ii)} creating an \textbf{adversarial} setting between two players (the generator and discriminator), and \textbf{iii)} harnessing \textbf{neural networks}. With regards to these three fundamentals, the state-of-the-art has leveraged \textbf{i} in terms of generating fixes to buggy code. It has also leveraged \textbf{iii} with a variety of neural models and designs. On the other hand, our approach has the novelty of leveraging \textbf{ii} on top of \textbf{i} and \textbf{iii}.

The adversarial property of GANs allows two neural networks (i.e. the generator and discriminator) to challenge, teach, and hence improve each other. As shown in this paper, our experimentation and evaluation of using GANs with two simple LSTM-based networks show promising results for APR. \emph{Our approach has the potential of improving all existing work by encapsulating them into our model}. For example, a model from the existing work (e.g., DLFix) can be used as the generator. By building the appropriate training environment, DLFix will get sharpened by, and sharpen, the discriminator. Furthermore, the discriminator can be designed in a variety of ways that are not limited to our chosen design. This is applicable in the same way to any model, or combination of models, from the state-of-the-art. For future work, we plan, and encourage the community, to research and report the results of building GAN frameworks that leverage the latest advances in the field of neural program repair. This includes, and is not limited to, targeting multi-line bugs and fixes and code change hunks as well as building adversarial environments between the existing approaches towards ultimately improving APR.

\subsection{Threats to Validity}

\subsubsection{External validity} Threats to external validity refer to the extent to which our findings can be generalized. In our study, one potential threat to external validity lies in the applicability of our results across a wide range of programming languages. Following established conventions in the field, we have conducted a thorough evaluation of our approach primarily using Java as the programming language. This evaluation involved comprehensive assessments against well-established benchmarks such as BigFix and Defects4J. While we did not examine the generalisability of our findings to other programming languages, we believe that our approach possesses the potential to be applied beyond Java as our methodology is general-purpose and not specific to Java. 

\subsubsection{Internal validity} Threats to internal validity refer to potential errors in our implementation and experiments. To mitigate this risk, we have carefully validated our implementation and experiments, ensuring their quality. Additionally, we have made our source code and results publicly accessible to the public. Therefore, we believe that the threat is minimal.

\subsubsection{Construct validity} Threats to construct validity refer to the suitability of our evaluation. To mitigate this risk, To minimize risks to construct validity, we conduct our empirical evaluation on widely-accepted benchmark datasets including BigFix and Defects4J. Hence, we believe the risk is minimal.

\section{Related Work}

\subsection{Generative Adversarial Networks for Software Engineering Tasks}
The emergence of Generative Adversarial Networks (GANs) has revolutionized numerous domains, enabling groundbreaking and influential applications in fields such as Computer Vision (CV)~\cite{goodfellow2014generative,wang2018transferring,collins2020editing,adler2018banach} and Natural Language Processing (NLP)~\cite{yu2017seqgan,guo2018long,pmlr-v137-kumar20a}. 
In the field of Software Engineering, GANs have also demonstrated remarkable potential, pushing the existing boundaries to new frontiers~\cite{zhang2018deeproad,porres2021online,guo2021towards,zhao2021guigan}.
Zhang et al.~\cite{zhang2018deeproad} introduced an autonomous driving systems testing technique that utilizes driving scenes generated by GANs. 
Porres et al.~\cite{porres2021online} and Gou et al.~\cite{guo2021towards} employed GANs to generate and validate test cases in software testing. 
Zhao et al.~\cite{zhao2021guigan} proposed GUIGAN, which employs various GAN models to automatically generate graphical user interface (GUI) designs. 
Motivated by these successful cases of generating realistic data, our proposed solution builds upon the GAN architecture for program repair.
However, different from prior approaches that directly employ original GANs to synthesize data, we propose a new GAN-based architecture that effectively addresses our specific problem.

\subsection{Learning-based Program Repair}



Over the past decade, significant progress has been made in the field of Automated Program Repair, leading to a substantial body of research~\cite{weimer2009automatically, kim2013automatic, qi2015analysis, le2016history, xuan2016nopol, long2016automatic, mechtaev2016angelix, xiong2017precise, le2017s3, mahajan2018automated, liu2019tbar, chen2019sequencer, le2021usability, ye2022neural, xia2022less, ye2022selfapr}. Among the various works, learning-based program repair techniques are particularly relevant to our study. The works primarily formulate APR as a Neural Machine Translation (NMT) problem: translating buggy code to fixed code using sequence-to-sequence (\emph{seq2seq}) models which typically consist of an \ul{encoder} and a \ul{decoder}. The formulation was pioneered by Tufano et al.\cite{tufano2018empirical} and Ratchet et al.\cite{hata2018learning}. These works differ in their targeted granularity levels, with the former focusing on method-level translation and the latter emphasizing statement-level translation. To enhance the effectiveness of the seq2seq model, Chen et al. proposed SequenceR\cite{chen2019sequencer}, which incorporates a copy mechanism.

Li et al. proposed, DLFix~\cite{li2020dlfix}, a seq2seq-based model consisting of two learning layers: a context learning layer and a transformation learning layer. DLFix captures the code structure by learning the Abstract Syntax Trees (ASTs) of the bug and its context but utilises a traditional decoder. In contrast, Recoder~\cite{zhu2021syntax} proposes a syntax-guided edit decoder with placeholder generation to address this limitation. Recoder employs a provider/decider neural architecture and specifically targets single-hunks. 

CoCoNuT~\cite{lutellier2020coconut} introduces Convolutional Neural Networks (CNNs) for Automated Program Repair (APR). CoCoNuT adopts an ensemble approach and utilises attention maps. This method is trained on large-scale data and requires significant computational resources and time. CURE~\cite{jiang2021cure} is another CNN-based model introduced later to address single-hunks, implementing a code-aware beam-search strategy.


Ye et al. introduced a novel approach called RewardRepair~\cite{ye2022neural} to address the overcorrection over different but reasonable fixes caused by cross-entropy loss. RewardRepair is a neural approach that employs execution-based back-propagation and incorporates compilation and test execution information during the training process. The authors observed that supervised models often lack project-specific and execution information, leading them to propose SelfAPR~\cite{ye2022selfapr}. SelfAPR is also an execution-based approach; however, it is self-supervised and utilizes test execution diagnostics to enhance performance.

In addition to the work mentioned above on patch generations, researchers have explored learning-based approaches for patch ranking and correctness assessment~\cite{long2016automatic, ye2021automated}. Prophet~\cite{long2016automatic} constructs a probabilistic model using historical bug fixes to rank correct patches. Rete~\cite{parasaram2023rete} utilizes program namespace representation to navigate the search space of patches effectively. ODS~\cite{ye2021automated} builds a classification model based on 4199 hand-crafted code features to identify patch correctness. Meanwhile, BERT-LR~\cite{tian2020evaluating} and Invalidator~\cite{le2023invalidator} utilize Large Language Models to automatically extract code features and evaluate patch correctness based on the learned features.


\section{Conclusions}

In summary, we have presented in this paper a novel deep learning approach named NEVERMORE to automate program repair. Our approach leverages the notion of Generative Adversarial Networks, which conveniently matches the adversarial nature of bugs and repairs, to generate program patches. Our approach is fully automated as it does not require hard-coding of bug-fixing rules. In addition, our solution does not require a set of test cases, which enables early intervention and repair of software bugs and avoids accumulating repair costs. Our approach does not perform an enumerative search for repairs but learns to fix a common class of bugs directly from patch examples. Thus, we offer a fast and scalable solution that can both enhance existing approaches and generalise across software applications. Our empirical evaluation suggests the feasibility of NEVERMORE and provides valuable insight for future research directions.

\bibliography{sn-bibliography}

\end{document}